\documentclass[preprint,12pt]{elsarticle}

\usepackage{amssymb}
\usepackage{graphicx}
\usepackage{subcaption}
\usepackage{bm}
\usepackage{color}
\usepackage{amsmath}
\usepackage{float}
\usepackage{gensymb}
\usepackage{array}
\usepackage{tabularx}
\usepackage{multirow}
\usepackage{changepage}
\journal{Journal of Nuclear Materials}

\begin{document}

\begin{frontmatter}

\title{Energy evolution in nanocrystalline iron driven by collision cascades} 

\author[inst1]{Ivan Tolkachev\corref{cor1}}
\ead{ivan.tolkachev@eng.ox.ac.uk}
\author[inst2]{Daniel Mason}
\author[inst2]{Max Boleininger}
\author[inst2]{Pui-Wai Ma}
\author[inst1]{Felix Hofmann\corref{cor2}}
\ead{felix.hofmann@eng.ox.ac.uk}
\cortext[cor1]{corresponding author}
\affiliation[inst1]{organization={Department of Engineering Science, University of Oxford},
            addressline={Parks Road}, 
            city={Oxford},
            state={Oxfordshire},
            postcode={OX1 3PJ}, 
            country={United Kingdom}}
\affiliation[inst2]{organization={United Kingdom Atomic Energy Authority},
            addressline={Culham Science Centre}, 
            city={Abingdon},
            state={Oxfordshire},
            postcode={OX14 3DB}, 
            country={United Kingdom}}

\begin{abstract}

Nanocrystalline materials are promising candidates for future fusion reactor applications, due to their high density of grain boundaries which may serve as sinks for irradiation induced defects. We use molecular dynamics to simulate collision cascades in nanocrystalline iron and compare these to collision cascades in initially defect free single crystals. We create nanocrystalline samples via Voronoi tessellation of initially randomly placed grain seeds and via severe plastic shearing. An irradiation induced annealing is observed whereby after $\sim$ 2 displacements per atom (dpa), irradiation drives all simulation cells to a single crystalline state. Irradiation-induced defects that distort the lattice generate elastic strain, so we use excess potential energy as a measure of defect content. At low doses, the Voronoi samples feature a few large, low energy grains, whereas the sheared samples show many small, high energy grains due to the high defect and grain boundary content caused by severe deformation. As dose increases beyond 1 dpa however, all nanocrystalline samples converge to a similar behaviour. Excess potential energy mirrors this trend, plateauing above $\sim$ 4 dpa. We hypothesise that the initially pristine cells will also reach a similar plateau after 5 dpa, which is seemingly confirmed by running a single instance of each cell type to 10 dpa. A model is developed to explain the energy evolution. We conclude that, regardless of initial structure, irradiation drives all cells toward the same energy state at high doses.

\end{abstract}




\end{frontmatter}


\section{Introduction}

High-energy neutrons are the main product of the deuterium-tritium nuclear fusion reaction \cite{KURTZ201951}. In operation, these neutrons will penetrate the reactor materials, interacting with the atomic nuclei. These interactions produce atomic recoils \cite{GILBERT2015121}, which create collision cascades within the material leading to the production of dislocations, voids, and other microstructural defects. These defects cause a degradation of the reactor materials' mechanical \cite{DUBINKO2021105522, SONG2024154998, SONG2024114144} and thermal \cite{REZA2022117926, HofmannThermal} properties.

Nanocrystalline materials have been considered for use in fusion applications \cite{BEYERLEIN2015125} due to their high grain boundary density, which may act as a sink for irradiation induced defects \cite{ElAtwaniSinks, BaiSinks, TscoppSinks, SamarasSinks}. These materials are characterised by their small grain size ($<$ 100 nm), and experimentally, a lower irradiation defect density has been observed for nanocrystalline materials when compared to coarse-grained material irradiated to the same dose \cite{NITA2004953, CHIMI2001355}.

Experimental techniques such as high-pressure-torsion, accumulative roll bonding, and equal channel angular pressing, have been extensively used to produce nanocrystalline materials through severe plastic deformation (SPD) \cite{AgarwalIOP, Borodachenkova17, ZHILYAEV2005277, ITO200932, STRANGWARDPRYCE2023101468}. Traditionally, molecular dynamics (MD) studies on nanocrystalline material have used Voronoi tessellation \cite{Voronoi1908} with randomly placed grain seeds to generate nanograined simulation cells. Recently, studies have been carried out on the creation of nanocrystalline materials in MD through SPD \cite{TolkachevSims, GuoFeComp}. Tolkachev \textit{et al.} \cite{TolkachevSims} induced grain refinement in iron by applying a high shear strain (up to $\gamma = 10$) to initially pristine simulation cells. Nanograin formation was directly observed and the grains were found to be stable at constant temperatures of 300 K, 500 K, and 800 K, for 1 ns.

Irradiation induced annealing processes have been observed in materials both experimentally \cite{SONG2024114144, NITA2004953, Atwater, Kaoumi, Bufford} and through MD simulations \cite{LeoMaPRM, LevoCascades, tolkachev2025atomistic}. Song \textit{et al.} \cite{SONG2024114144} irradiated nanocrystalline and coarse-grained Eurofer97 with self-ions. An irradiation induced annealing process was revealed through X-ray diffraction measurements which showed grain coarsening and a reduction of dislocation density with increasing irradiation dose. Ma \textit{et al.} \cite{LeoMaPRM} used MD simulations to study irradiation damage in nanocrystalline tungsten by utilising the creation relaxation algorithm \cite{DudarevCRA}. Similarly to the experimental findings, a grain coarsening occurred with increasing dose. Levo \textit{et al.} \cite{LevoCascades} also observed irradiation induced grain growth in nanocrystalline nickel subjected to collision cascade simulations. Tolkachev \textit{et al.} \cite{tolkachev2025atomistic} reported a grain coarsening with irradiation in nanocrystalline iron. This combined atomistic and experimental study also confirmed grain coarsening in self ion-irradiated, nanocrystalline iron discs, and the MD data suggested a linear grain growth, which was consistent with the experimental values. 

Every grain within a material will have inherent elastic energy due to the elastic strain field associated with grain boundaries and crystal defects. It is expected that the irradiation induced annealing process reported both in experiments \cite{SONG2024114144, NITA2004953, Atwater, Kaoumi, Bufford} and MD simulations \cite{LeoMaPRM, LevoCascades} will reduce the elastic energy, as grain growth and annealing occurs. The potential energy will be a sum of defect formation energy, grain boundary energy, and elastic energy. As such, tracking the excess potential energy as a function of irradiation is a useful method of measuring the integrated effect of grain growth and defect reduction for initially nanocrystalline MD simulation cells.

Reduced activation ferritic/martensitic (RAFM) steels such as Eurofer97 and T91 are promising candidate materials for future nuclear fusion reactors. Previous studies have explored their irradiation resistance \cite{Garner_JNM_2000, Boutard, CABET2019510, MAZZONE2017655}, and Garner \textit{et al.} \cite{Garner_JNM_2000} reported that RAFM steels with body-centred-cubic structure, were more resistant to irradiation induced swelling than austenitic steels. RAFM steels are iron-based alloys and in this work, we use iron (Fe) as a prototypical material to gain mechanistic insight into its evolution under irradiation.

An established method for simulating irradiation damage using MD simulations is through collision cascades \cite{LevoCascades, BoleiningerCascade, GRANBERG2020151843, SAND201864, GranbergCascades}. This method selects random atoms in a simulation cell and assigns velocity in a random direction corresponding to an expected primary-knock-on atom (PKA) energy. 

In this work, nanocrystalline iron cells created through Voronoi tessellation \cite{Voronoi1908} and through the shearing method reported by Tolkachev \textit{et al.} \cite{TolkachevSims} are considered. These cells are subjected to irradiation through collision cascade simulations and the evolution of excess potential energy within the cells is tracked as a function of dose. First, the excess potential energy is calculated for all cells and compared. Then, the excess potential energy is calculated for each individual grain for different types of starting microstructure. This gives insight into how grains evolve during irradiation. Finally, the implication of these results is discussed before drawing conclusions.

\section{Methods}

\subsection{Simulation Setup}

LAMMPS was used for all MD simulations \cite{LAMMPS} in this work. Two distinct methods were employed to generate nanocrystalline iron cells; Voronoi tessellation \cite{Voronoi1908} and the high shear strain method proposed by Tolkachev \textit{et al.}\cite{TolkachevSims}, hereby denoted as the \textit{sheared} cells. In the case of the Voronoi cells, two different grain numbers were seeded for comparison: one with 5 grains per cell (d$_{c}$ = 16.6 nm) and the other with 20 grains per cell (d$_{c}$ = 5.2 nm). The diameters are calculated using the spherical grain approximation. Five independent cells are created for each initial condition, and the results and errors below are calculated with respect to five simulations for each type of microstructure. For comparison, five additional initially pristine iron cells were also irradiated through collision cascades.

All cells contained 1,024,000 atoms which corresponds to 80 x 80 x 80 unit cells of BCC iron. The cell relaxation method reported by Ma \textit{et al.}\cite{LeoMaPRM} was followed for all distinct simulation cells. An extra step was added to the sheared cells in that they were initially thermalised to 300 K and sheared up to $\gamma = 10$. By utilising the method in \cite{LeoMaPRM}, the cells achieved stress free conditions and were brought down to a temperature of 0 K before the collision cascade simulations. The dataset considered here was also used in Tolkachev \textit{et al.} \cite{tolkachev2025atomistic}. Previously, Tolkachev \textit{et al.} determined that the Voronoi cells' grain volume correlated well to the equation of Ference \textit{et al.} \cite{FERENC2007518}, whilst the grain volume distribution of the sheared cells was fit to a Zipf-Mandelbrot model \cite{ZipfMandlebrot}. 

The potential used in this study is the Mendelev \textit{et al.}\cite{MENDELEVPOT} Fe2 potential. This belongs to the Ackland-Mendelev family of iron potentials which have been used extensively for the study of irradiation damage effects in iron and iron alloys \cite{SAND201864, DudarevCRA, KEDHARNATH2019444, YE2021152909, MALERBA201019}. Indeed, Malerba \textit{et al.} \cite{MALERBA201019} commented that the Mendelev-type potentials are the best choice for studying irradiation effects on iron when compared to density functional theory. 

Collision cascade simulations were employed to replicate neutron damage in the nanocrystalline iron. The method in Tolkachev \textit{et al.} \cite{tolkachev2025atomistic} was followed which is based on the collision cascade method developed by Boleininger \textit{et al.} \cite{BoleiningerCascade}. The Stopping and Range of Ions in Matter (SRIM) \cite{ZIEGLER20041027} code was used to determine the recoil energies of primary knock on (PKA) atoms in iron. These recoil energies were then sampled from for each collision cascade simulation to ensure a target dose was obtained in each step. Appendix A of Tolkachev \textit{et al.} \cite{tolkachev2025atomistic} shows the recoil energy distribution that was sampled in this work. It is dominated by lower energy recoils (30 - 200 eV), with a much lower probability of selecting higher energy recoils (200 eV - 20 keV). The exclusion radius of each PKA is also calculated to ensure there is no overlap between cascades.

Each distinct simulation cell was irradiated to 5 dpa, with a single instance of each cell type being irradiated to 10 dpa through collision cascade simulations. Each primary-knock-on (PKA) atom energy was converted to a damage energy, through the Lindhard \cite{Lindhard} model, which informed the Norgett-Robinson-Torrens (NRT-dpa) model \cite{NRT1,NRT2} used to determine the number of atomic displacements generated by any given cascade. This method is used to measure the applied dpa between steps. 

High dose rates ($d\phi/dt \sim 1 \ dpa/50 \ ns$) are present in this simulation due to the time limitations of MD simulations, effectively meaning that the MD simulations will only show fast processes, such as Stage I and II recovery, but slow processes with high thermal barriers will be missed. Hence, the cascade simulations are propagated in the athermal regime. This has previously produced a credible representation of irradiated material \cite{BoleiningerCascade} and is also suitable for comparison with room temperature experimental data on iron.

\subsection{Computational Analysis}

The OVITO code \cite{Stukowski_2010} was also used for the analysis. Polyhedral template matching (PTM) was used to determine local atomic orientations \cite{Larsen_2016}, which were fed into the grain segmentation modifier to identify which atoms were part of which grains. To calculate the mean excess potential energy per grain, the excess potential energy of each atom in the grain is calculated, and a mean over all the atoms is taken. The excess potential energy of the entire simulation cell is also calculated using LAMMPS. Excess is taken to mean the potential energy, in eV/atom, of a given grain or cell subtracted from the potential energy of a pristine, defect free iron cell, at room temperature with traction free boundary conditions.

\section{Results}

\begin{figure}
\centering
\includegraphics[width=\linewidth]{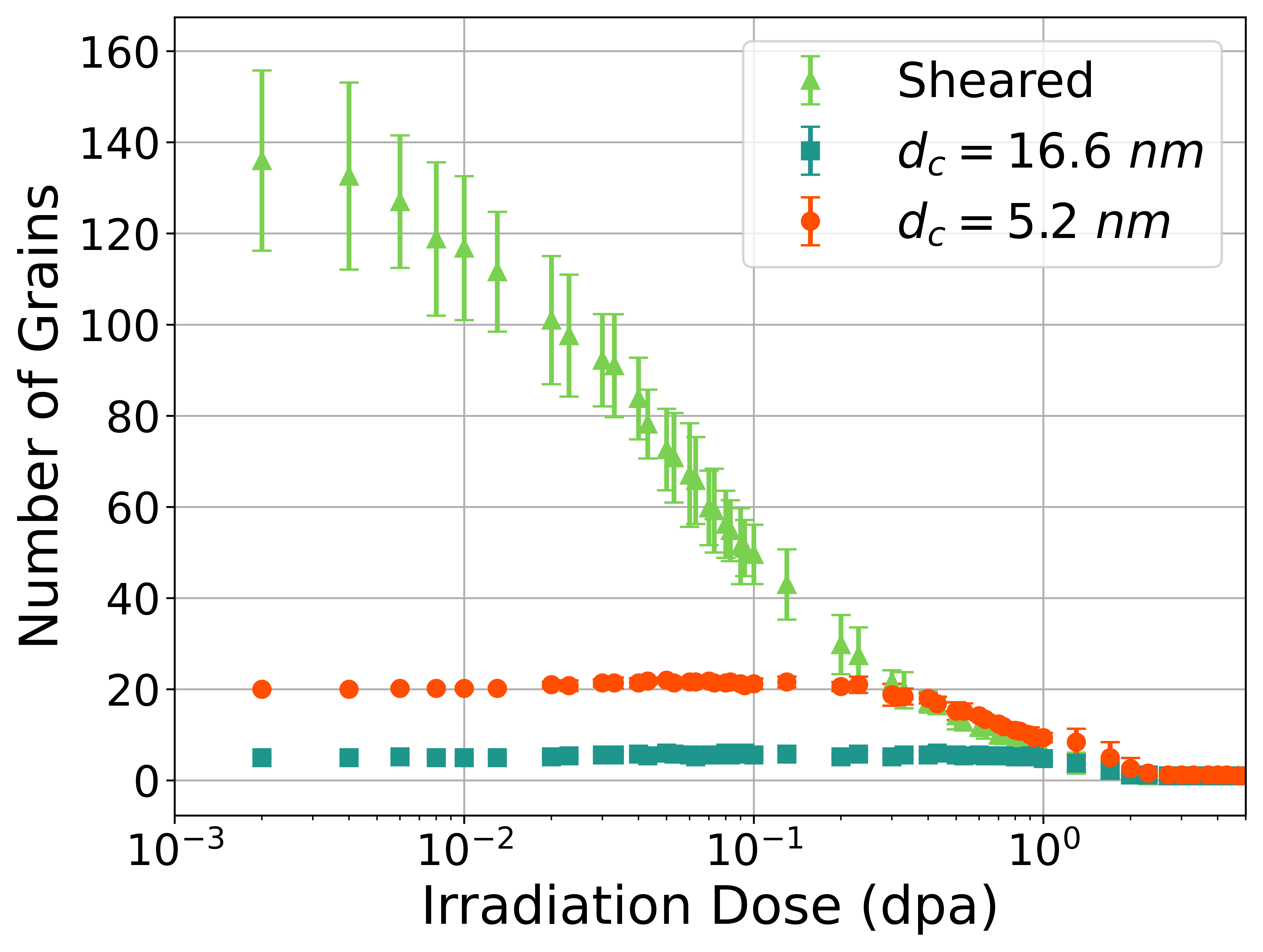}
\caption{Grain number, obtained using grain segmentation in Ovito, as a function of dose for all initially nanocrystalline cells. A grain was taken to be a minimum of 100 atoms.} \label{fig:Fig1}
\end{figure}

Figure \ref{fig:Fig1} shows the mean grain number as a function of dose for all initially nanocrystalline iron cells, averaged over the five distinct simulations of each type of starting microstructure. A clear irradiation induced annealling process is shown whereby the grains in all the initially nanocrystalline cells grow and become single crystalline after $\sim$ 2 dpa. In Tolkachev \textit{et al.} \cite{tolkachev2025atomistic}, which uses the same dataset, the irradiation induced grain growth was found to be consistent with random grain growth. Furthermore, the study \cite{tolkachev2025atomistic} confirmed irradiation induced grain growth experimentally in self-ion irradiated, nanocrystalline iron. 

Similar behaviour has previously been observed using molecular dynamics simulations \cite{LeoMaPRM, LevoCascades}. Ma \textit{et al.} \cite{LeoMaPRM} observed irradiation induced grain growth in nanocrystalline tungsten, whilst Levo \textit{et al.} \cite{LevoCascades} showed a grain growth in nanocrystalline nickel, with all cells becoming single crystalline with increased dose. Experimentally, irradiation induced annealling has also been observed in numerous studies \cite{SONG2024114144,NITA2004953, Atwater, Kaoumi, Bufford}. 

\begin{figure}
\centering
\begin{subfigure}{0.78\textwidth}
\includegraphics[width=\linewidth]{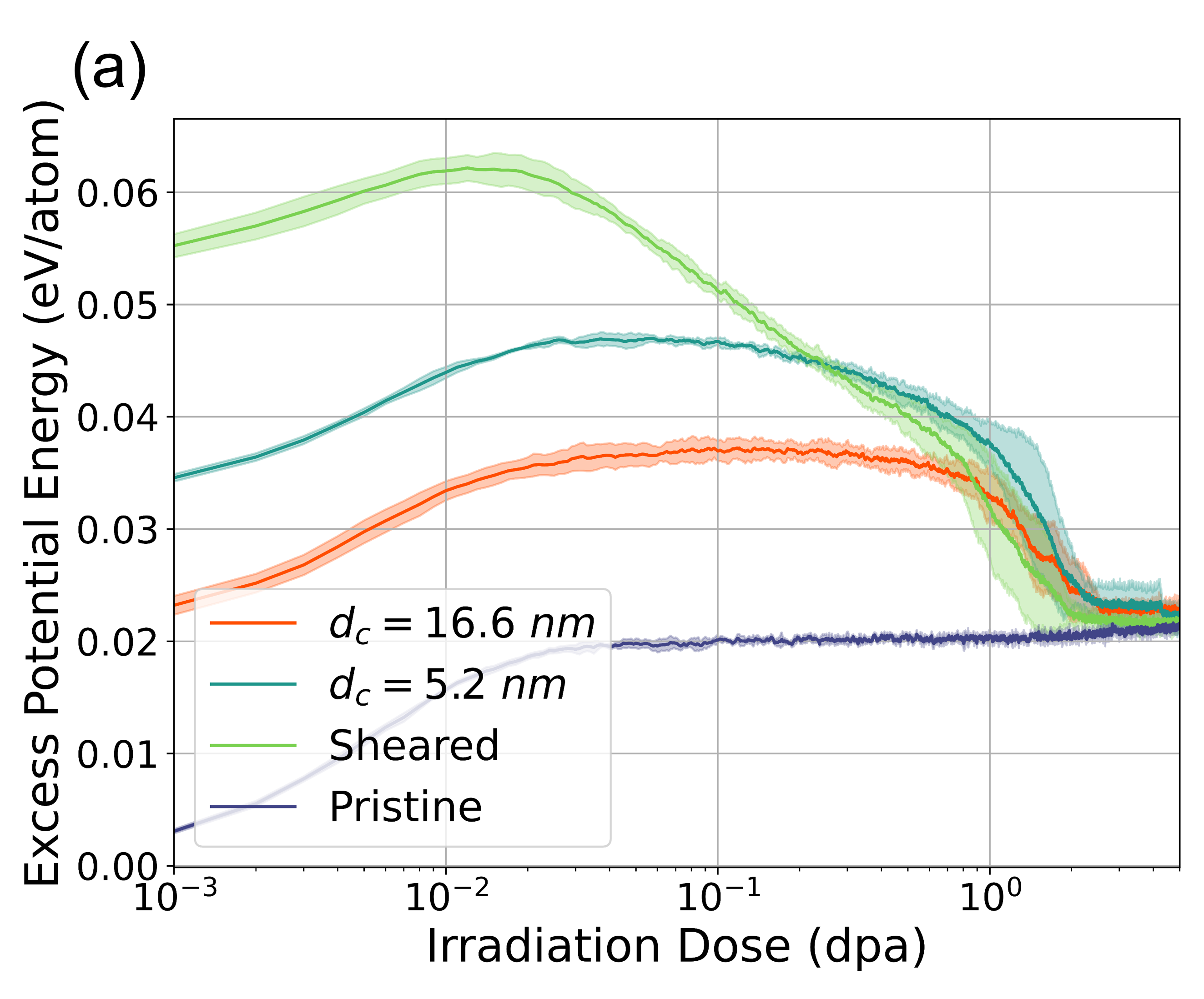}
\label{fig:Fig2a}
\end{subfigure}

\begin{subfigure}{0.78\textwidth}
\includegraphics[width=\linewidth]{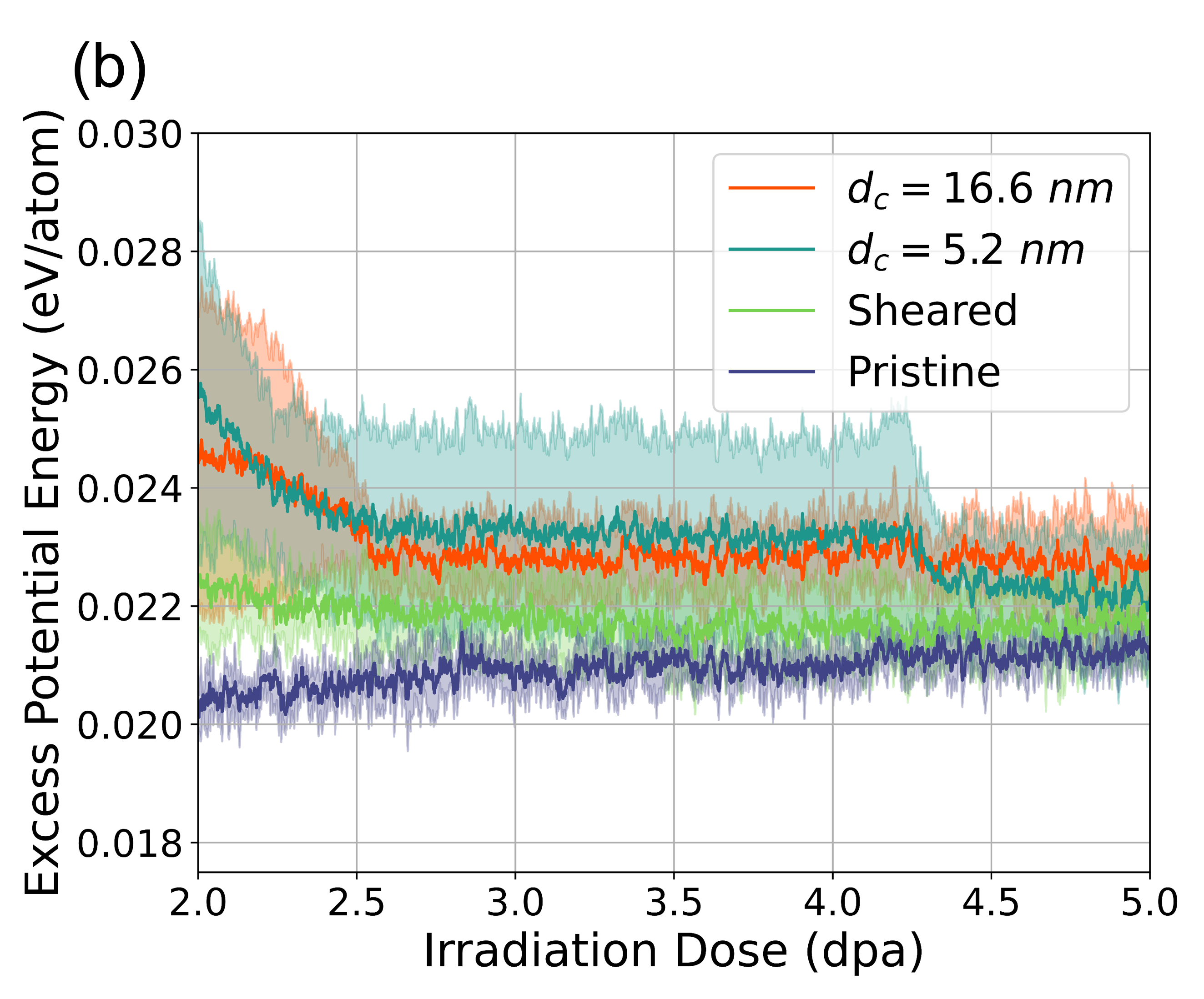}
\label{fig:Fig2b}
\end{subfigure}

\caption{Excess potential energy versus dose for initially nanocrystalline and pristine cells in eV/atom; (a) Plotted between 10$^{-3}$ dpa and 5 dpa, (b) Plotted between 2 dpa and 5 dpa.} \label{fig:Fig2}
\end{figure}

Excess potential energy versus dose is plotted in Figure \ref{fig:Fig2}. The initially pristine cells are also added here for comparison. Figure \ref{fig:Fig2}(a) shows the overall excess potential energy, in eV/atom, for each starting configuration averaged over five independent simulations. As expected, at 0.001 dpa, the sheared cells have the highest excess energy due to the large number of grain boundaries and defects present within the cells. The d$_{c}$ = 5.2 nm cells have the next highest, followed by the d$_{c}$ = 16.6 nm cells, which is again expected as having smaller grains and more grain boundaries will raise the excess potential energy.

All initially nanocrystalline cells follow a similar trend; excess potential energy increases before decreasing and seemingly plateauing at similar values above $\sim$ 4 dpa. This is more evident in Figure \ref{fig:Fig2}(b) which shows a zoomed in view of excess potential energy between 2 and 5 dpa. It is noticeable, particularly after 4.5 dpa, that the excess potential energy in the initially nanocrystalline cells seemingly converges at similar values of $\sim$ 0.022 - 0.023 eV/atom, with much overlap in the standard deviations. This seems to suggest that, irrespective of the initial starting configuration, all initially nanocrystalline cells will converge at similar values of excess potential energy at high doses.

In Figure \ref{fig:Fig2}(a), the initially pristine cells do not follow the trend of the initially nanocrystalline cells, and the excess potential energy only increases with dose. Interestingly, the excess potential energy of the pristine cells nearly matches that of the initially nanocrystalline cells, being $\sim$ 0.021 eV/atom at 5 dpa. We also note that the excess potential energy in the pristine cells is seemingly still increasing at 5 dpa and hypothesise that it will continue to increase until it reaches the range of the initially nanocrystalline cells.

\begin{figure}
\centering
\includegraphics[width=\linewidth]{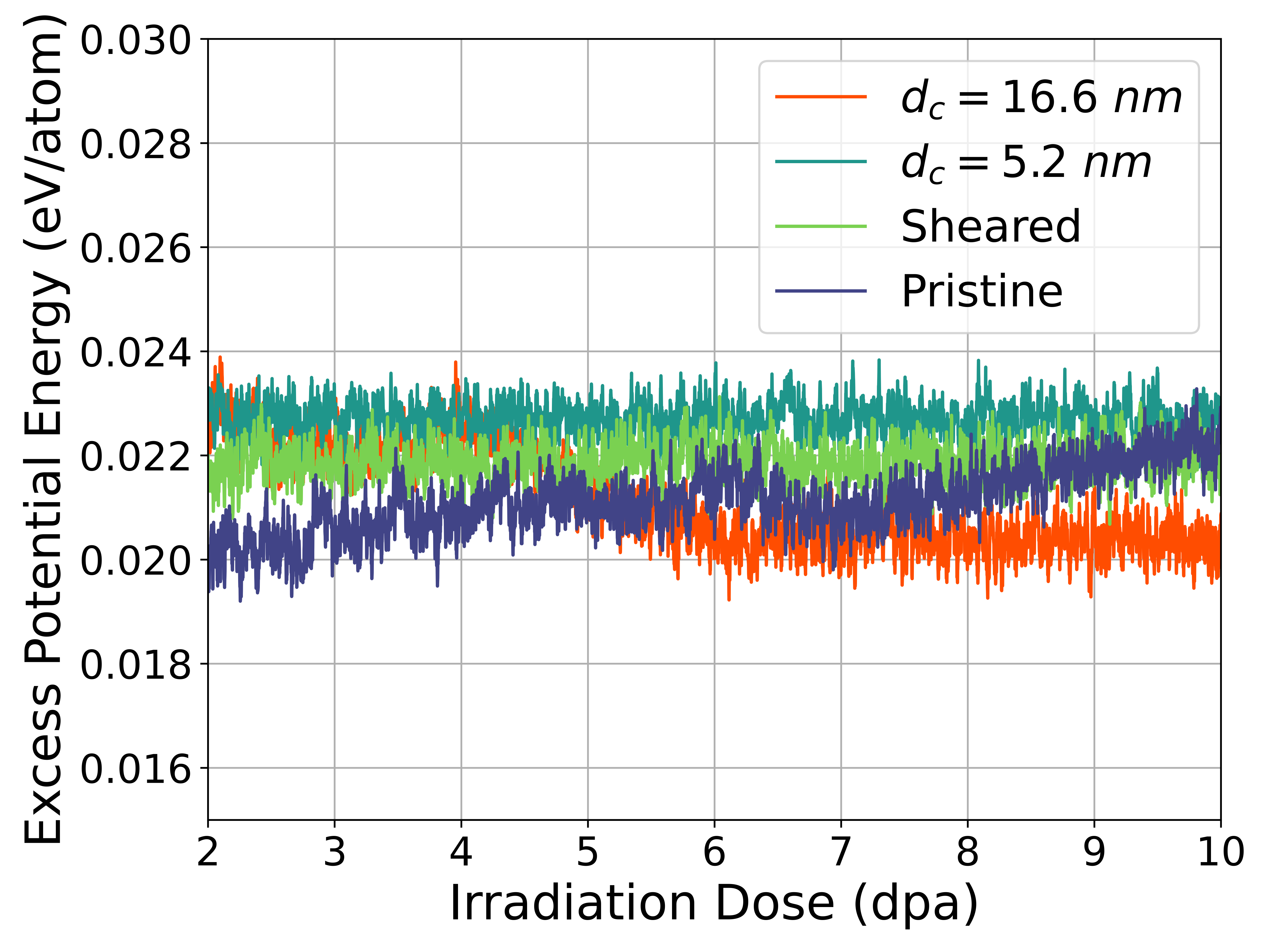}
\caption{Damage against excess potential energy for a single instance of each simulation type between 2 and 10 dpa.} \label{fig:Fig3}
\end{figure}

To test this hypothesis, a single instance of each simulation type was run to a damage of 10 dpa and the excess potential energy for these simulations, between 2 and 10 dpa, is shown in Figure \ref{fig:Fig3}. It is noticeable that both the sheared and d$_{c}$ = 5.2 nm cells have plateaued beyond 2 dpa at $\sim$ 0.022 eV/atom. However, the pristine cells is at 0.020 eV/atom at 2 dpa, and rises to 0.022 eV/atom by 9.5 dpa, seemingly plateauing after this dose. This seems to confirm the hypothesis above, whereby all simulation cell types will converge to a similar excess potential energy at high dose, irrespective of their initial starting configuration. Figure \ref{fig:Fig3} does however show that the excess potential energy for the d$_{c}$ = 16.6 nm cell reduces to $\sim$ 0.020 eV/atom after 6 dpa and plateaus here. It is not immediately obvious why this occurs however, this variation is on the same scale as simulation-to-simulation fluctuations.

\begin{figure}
  \centering
  \begin{adjustwidth}{-0.2\textwidth}{-0.2\textwidth}
    \includegraphics[width=\linewidth]{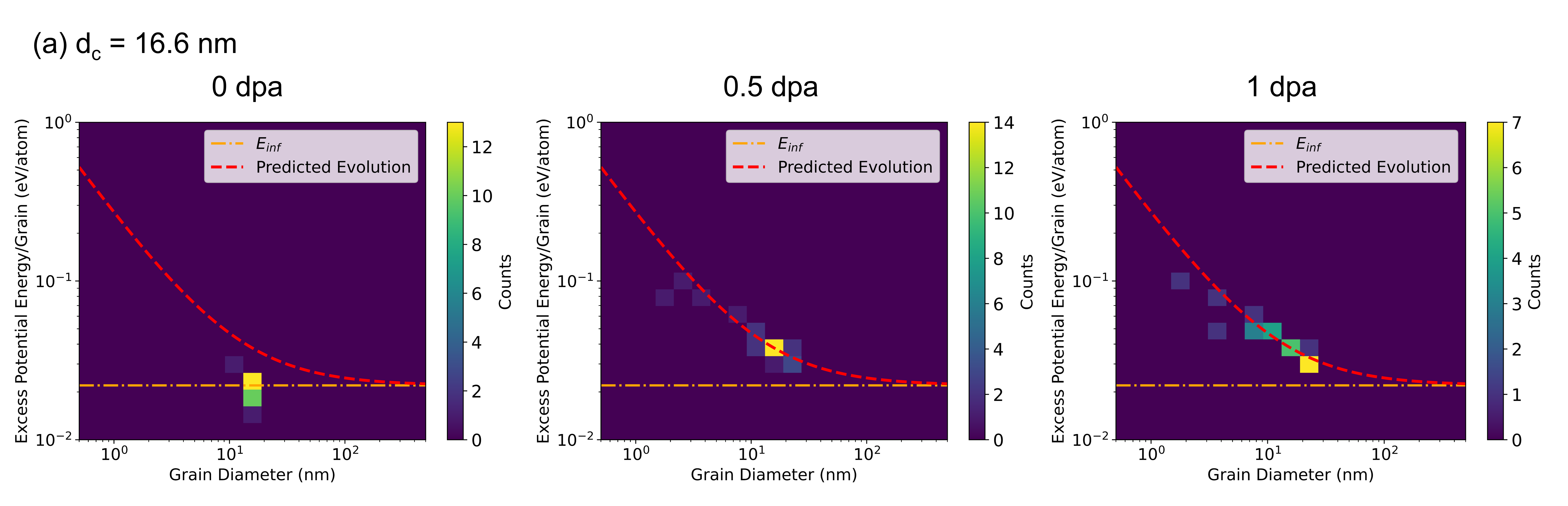}

    \includegraphics[width=\linewidth]{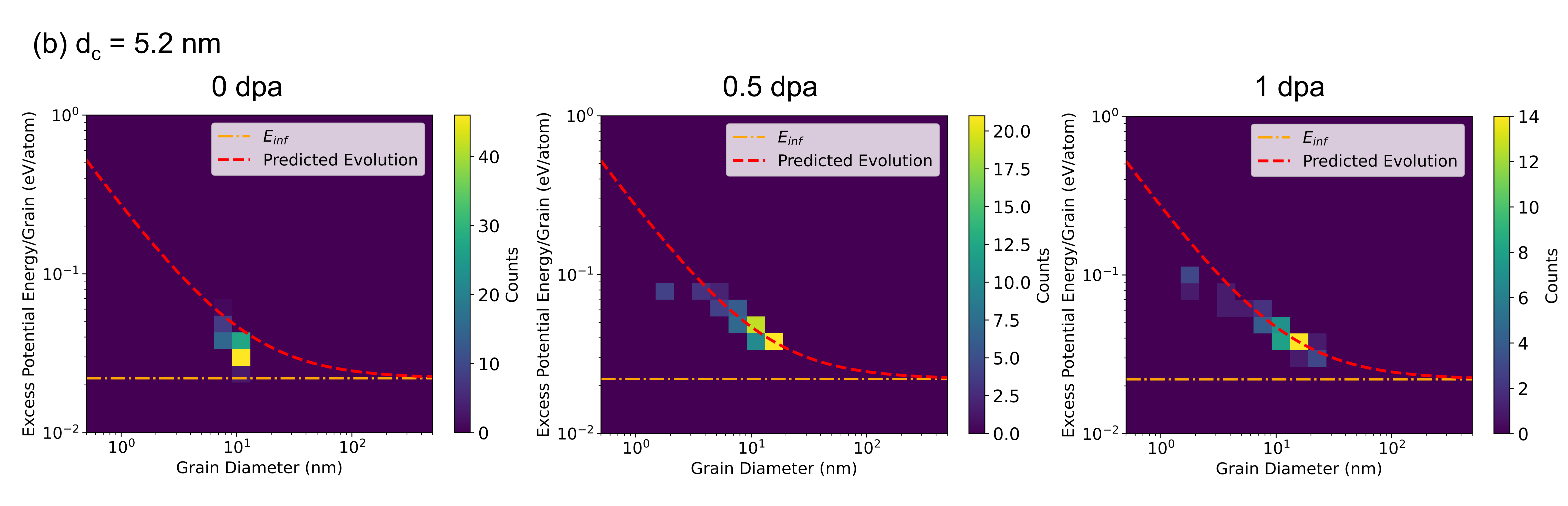}

    \includegraphics[width=\linewidth]{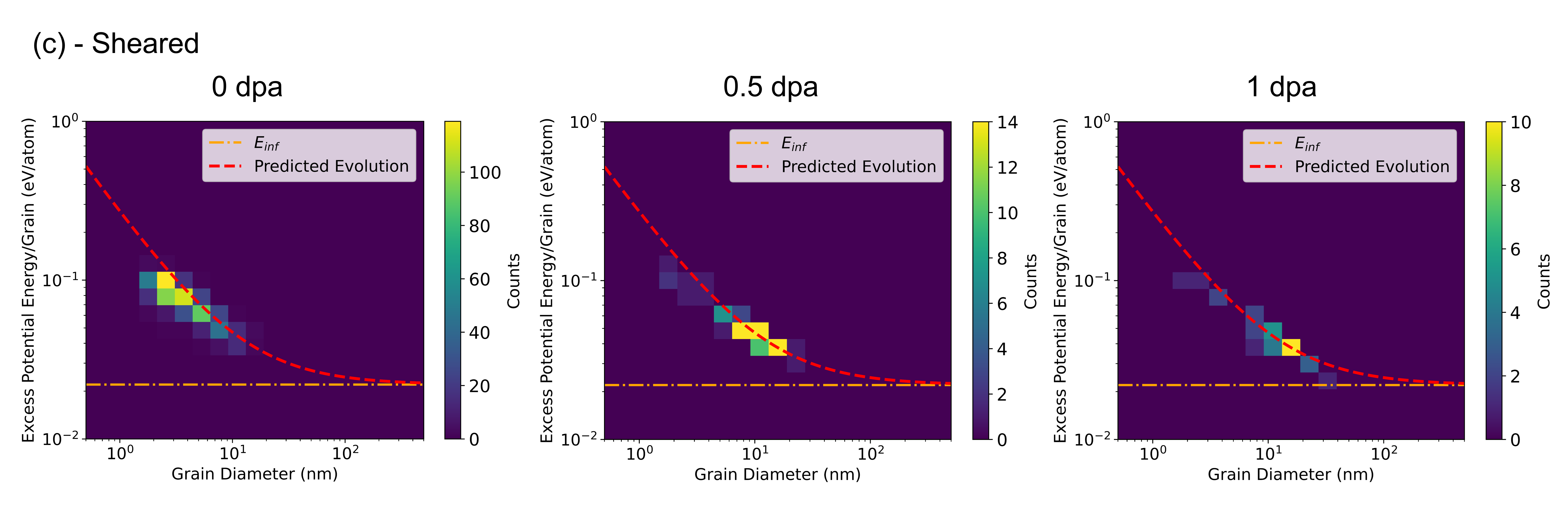}
  \end{adjustwidth}

  \caption{Excess potential energy histograms for individual grains for each starting nanocrystalline configuration; (a) d$_{c}$ = 16.6 nm, (b) d$_{c}$ = 5.2 nm, and (c) sheared cells.}
  \label{fig:Fig4}
\end{figure}

To evaluate how the excess potential energy evolves for individual grains, histograms of excess energy versus grain size for all simulations are presented in Figure \ref{fig:Fig4}. In Figure \ref{fig:Fig4}(a), the data for the d$_{c}$ = 16.6 nm cells is shown. Initially, at 0 dpa, all grains have similar excess energy and size. By 0.5 dpa, there is a noticeable grain growth, with some grains getting larger. Note here, that some grains get smaller, corresponding to a higher excess energy. By 1 dpa, there are more larger grains of lower energy and some smaller grains of higher energy. Figure \ref{fig:Fig4}(b) shows the same data for the d$_{c}$ = 5.2 nm cells. Again, a similar grain growth process is observed, and a direct correlation between grain size and excess potential energy is visible. At 1 dpa, there are numerous larger grains with lower energy and smaller grains with higher energy, as with the d$_{c}$ = 16.6 nm cells.

The data for the sheared cells in Figure \ref{fig:Fig4}(c) follows a different evolution. As in Figure \ref{fig:Fig1}, there are many more grains in the sheared cells, and this is reflected in the 0 dpa figure. By 0.5 dpa, the number of large grains has increased, with some smaller grains of higher energy, and a noticeable decrease in grain number as per Figure \ref{fig:Fig1}. Note the similarity in histograms between all cell types at 1 dpa. Whilst the number of grains is different, there is a noticeable trend in evolution notwithstanding the starting microstructure.

Appendix A depicts the histogram data for all cells at 5 dpa, and this shows that the excess potential energy per grain does plateau at $E_{inf}$. At this dose, all cells are single crystalline so the grain diameter becomes arbitrary.

\section{Discussion}

The molecular dynamics simulations presented in this work demonstrate a direct irradiation induced grain growth process as shown in Figure \ref{fig:Fig1}. This figure shows that the initially nanocrystalline cells experience a grain coarsening, and all become single crystalline after $\sim$ 2 dpa. This is accompanied by a reduction in excess potential energy within grains as shown in Figure \ref{fig:Fig4}. In Figures \ref{fig:Fig4}(a) and \ref{fig:Fig4}(b), the Voronoi-generated cells start off with evenly sized grains of lower energy. At 0.5 dpa, some grains get larger, whilst some get smaller, but its noticeable that the excess potential energy in grains decreases with size increase.

Figure \ref{fig:Fig4}(c) shows the excess potential energy for the grains within the sheared cells at increasing doses. It is evident that the starting energies for these cells are very different to those of the Voronoi-generated cells. In fact, there are many small grains at 0 dpa with high energies, and this is also reflected in Figure \ref{fig:Fig1} which shows the presence of many small grains in the sheared cells. Due to the nature of the cell creation, the grains within the sheared cells are expected to have higher energy as they have inherent material defects such as dislocations and interstitials as a result of the plastic deformation process. By 0.5 dpa, the irradiation induced annealing process leads to grain growth and the excess potential energy in most grains reduces as grain boundaries and other defects are annealed.

Interestingly, at 1 dpa, all histograms show that, irrespective of the initial configuration, all grains converge to a similar excess potential energy. This is also shown in Figure \ref{fig:Fig2}(a) which shows the excess potential energy for all cell types, averaged over five distinct simulations. Whilst the initially nanocrystalline cells behave differently at low doses, it is noticeable that their excess potential energy converges at higher doses. 

Figure \ref{fig:Fig2}(b) shows that both Voronoi-generated and sheared cells converge to $\sim$ 0.022 - 0.023 eV/atom by $\sim$ 4.5 dpa, with overlapping standard deviations across all five simulation types. Pristine cells, by contrast, exhibit a continuous rise in excess potential energy with dose. Extending one simulation of each cell type to 10 dpa (Figure \ref{fig:Fig3}) confirms that the pristine cell’s energy keeps increasing until $\sim$ 9.5 dpa, then plateaus at $\sim$ 0.022 eV/atom.

To explain this behaviour, we develop a hypothesis on how the excess potential energy in the grains is dependent on grain size, when driven by collision cascades. It is proposed, based on the observable data in Figure \ref{fig:Fig4}, that when grains are smaller, the size of the grain dominates the excess potential energy within that grain giving:

\begin{equation} \label{eq1}
    E = Ad^{-\phi}
\end{equation}
\\

where $E$ is excess potential energy density, $-\phi$ is a constant governing decay, $d$ is grain size, and $A$ is a constant of proportionality. It is noted from Figure \ref{fig:Fig4} that at small grain size, the energy evolution looks linear. If we assume that the excess potential energy is dominated by grain boundary energy, then the energy per unit volume roughly scales with grain boundary energy/volume giving $\sim 1/d$. Hence, the value of $\phi = -1$. 

We also propose that at some high dose, the size of the grains becomes irrelevant and that the radiation damage dominates the excess potential energy value instead. This is shown as $E_{inf}$ in Figure \ref{fig:Fig4}. $E_{inf}$ corresponds to the excess potential energy caused by radiation damage in the high dose limit, and is calculated by determining the excess potential energy in the pristine cell at 10 dpa in Figure \ref{fig:Fig3}, which was found to be $2.2 \times10^{-2}$ eV/atom.

This term is added as it is noticeable in Figure \ref{fig:Fig4}, and in the 5 dpa figures in  Appendix A, that the excess potential energy per grain does plateau at $E_{inf}$. At this dose, all cells are single crystalline so the grain diameter becomes arbitrary. Hence, the predicted evolution becomes:

\begin{equation} \label{eq2}
    E = Ad^{-1} + E_{inf}
\end{equation}
\\

\begin{figure}
\centering
\includegraphics[width=\linewidth]{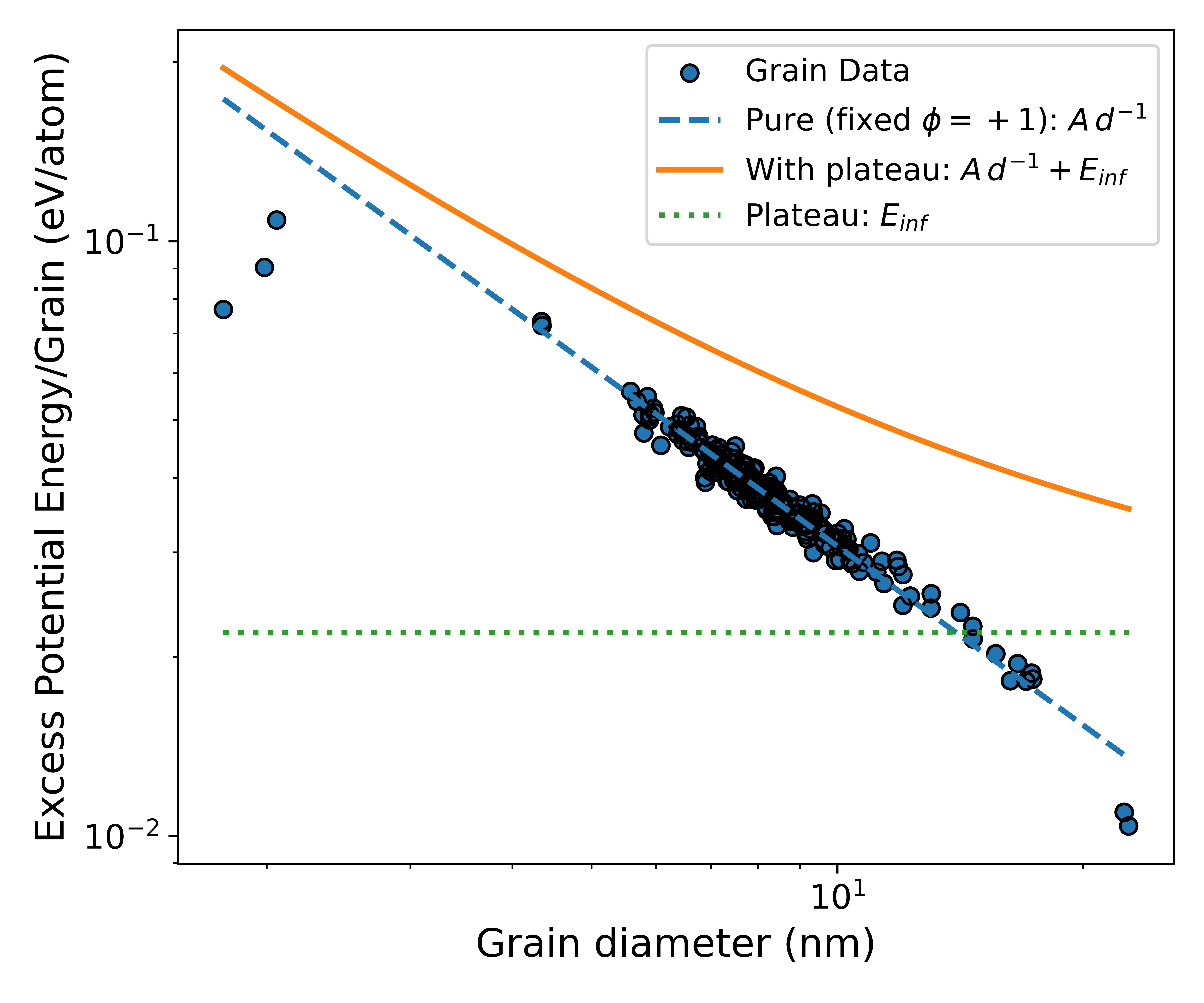}
\caption{Predicted evolution of excess potential energy with increasing grain size fitting.} \label{fig:Fig5}
\end{figure}

The value of $A$ was determined to be $0.307 \pm 8.972\times10^{-4}$ eVnm when fitting and keeping $\phi = -1$. The fitting method is shown in Figure \ref{fig:Fig5}. The pure $Ad^{-1}$ part is found by fitting data to the energy of grains of various sizes generated through Voronoi tessellation and relaxed. These represent perfect, defect free grains without the involvement of radiation damage. The extra $E_{inf}$ term captures the added effect of irradiation damage in the limit of an infinite crystal, as discussed above. Nonetheless, when plotting Figure \ref{fig:Fig4}, and carrying out the fitting in Figure \ref{fig:Fig5}, the values of $\phi$ was set to $-1$. The value of $A$ was analytically determined in Appendix B, and was found to be in the range of $A = 0.219 - 0.489$ eVnm, which coincides well with the fitted value of $A = 0.307$ eVnm, giving confidence to the fitting.

It is noticeable, from Figure \ref{fig:Fig4}, that the model is able to predict the observed excess potential energy evolution in the grains of the nanocrystalline material. Appendix A shows that, at 5 dpa, all grains plateau at the value of $E_{inf}$. Note, that the grain here becomes infinitely large as the MD simulation cells become single crystalline after $\sim$ 2 dpa, as shown in Figure \ref{fig:Fig1}.

Our results suggest that, from an energetic perspective, all initially nanocrystalline cells will converge to similar excess potential energy at high doses, notwithstanding their differing initial configurations and subsequent evolution at low doses. Furthermore, it appears that an initially pristine cell will also reach the same energetic state as the initially nanocrystalline cells in the high dose limit.

\section{Conclusion}

We have studied the energy evolution of both initially nanocrystalline and pristine cells subjected to cascade damage using MD simulations. The nanocrystalline cells were created through both Voronoi tessellation and plastic shearing deformation. Subsequently, our results allow the following conclusions:

\begin{itemize}
    \item An irradiation induced annealing process was observed whereby all initially nanocrystalline cells became single crystalline after $\sim$ 2 dpa.
    \item The annealing was further confirmed by considering the excess potential energy in individual grains for the initially nanocrystalline iron cells. All cell types showed that larger grains tended to have lower excess potential energy than smaller grains.
    \item A model was derived to explain the evolution of excess potential energy within the grains with irradiation as some grains increased in size and some became smaller. This was found to correlate well with the observed energy data for the grains in the MD simulation cells.
    \item This trend is further shown when looking at the excess potential energy in the cells. Whilst at low doses, all initially nanocrystalline cells have differing evolution, at doses above $\sim$ 4 dpa, the excess potential energy in all cells seemingly converges. Interestingly, the pristine cells also reach a similar value of excess potential energy at 5 dpa, albeit slightly lower. It was noticeable that the excess potential energy in the initially pristine cells was still increasing at 5 dpa. We hypothesised that the energy would continue increasing at higher doses to the same level of the initially nanocrystalline cells. A single instance of each cell type was run to a dose of 10 dpa to test this hypothesis. This data showed that the pristine cell's excess potential energy plateaued at the same point as some of the initially nanocrystalline cells after $\sim$ 9.5 dpa.
    \item All data seems to point to the fact that, irrespective of initial cell configuration, from an energetic perspective, all cells will converge at near identical values at high doses.

\end{itemize}

\section{Data Availability}

All input scripts and simulation data presented in the current work are available at \textit{A link will be provided after the review process and before publication.} 

\section*{Acknowledgements}

The authors gratefully acknowledge the Department of Engineering Science at the University of Oxford for their contribution to the funding of the project. This work has been carried out within the framework of the EUROfusion Consortium, funded by the European Union via the Euratom Research and Training Programme (Grant Agreement No 101052200 — EUROfusion) and from the EPSRC [grant number EP/W006839/1]. To obtain further information on the data and models underlying this paper please contact PublicationsManager@ukaea.uk. Views and opinions expressed are however those of the author(s) only and do not necessarily reflect those of the European Union or the European Commission. Neither the European Union nor the European Commission can be held responsible for them. The authors gratefully acknowledge the use of the ARCHER2 UK National Supercomputing Service (https://www.archer2.ac.uk) under project e804 and associated support services provided by the ARCHER2 Service Desk in the completion of this work. This work used the Cambridge Service for Data Driven Discovery (CSD3) Service (www.csd3.cam.ac.uk).

\appendix
\section{Histogram Data at 5 dpa}

\begin{figure}
\centering
\begin{adjustwidth}{-0.2\textwidth}{-0.2\textwidth}
    \includegraphics[width=\linewidth]{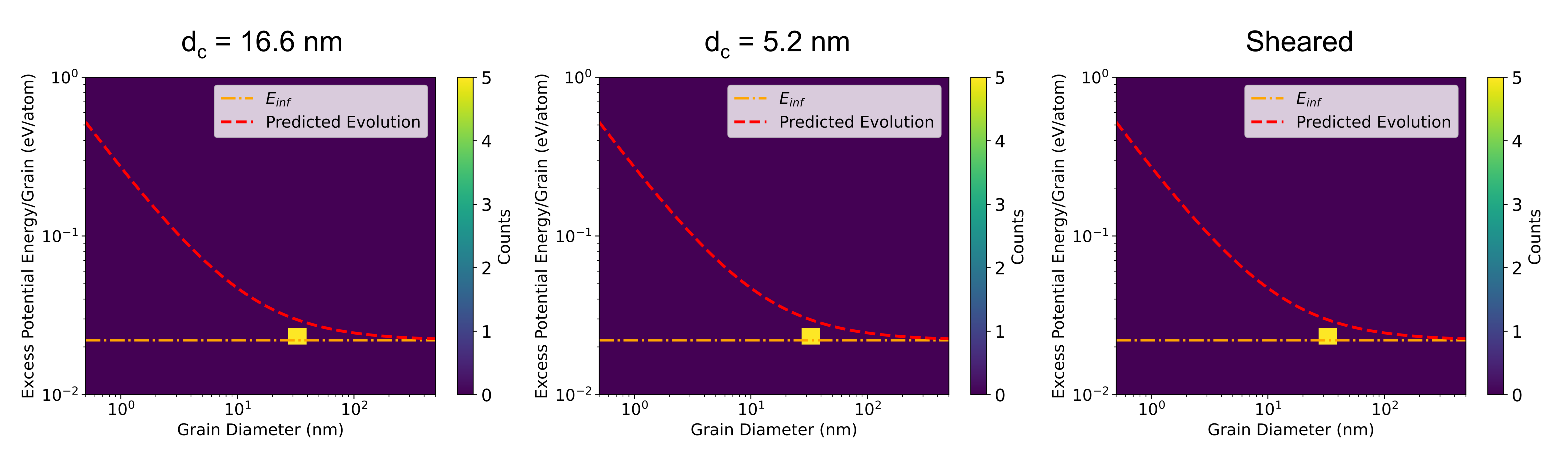}
\end{adjustwidth}
\caption{Histograms of excess potential energy versus grain diameter at 5 dpa for all initial starting configurations. By this dose, all cells are single crystalline which means that grain diameter is the same as the cell size, corresponding to the single crystal limit.} \label{fig:AppA}
\end{figure}

Figure \ref{fig:AppA} depicts the histograms of excess potential energy in the grains at 5 dpa for each initial simulation cell condition. At this dose, all grains are single crystalline and due to periodic boundary conditions, grains diameter becomes arbitrary. Nonetheless, the box size is 28 nm x 28 nm x 28 nm, and hence, this shows as the grain diameter. The histograms show that at high dose, the initially nanocrystalline cells all converge on the value of $E_{inf}$.

\section{Analytical Analysis for Proportionality Constant}

As mentioned before, the fitted value of $A$ was found to be $0.307 \pm 8.972\times10^{-4}$ eVnm. To determine whether this value is suitable, an analytical analysis was carried out as below.

Firstly, we start with the equation $ E = Ad^{-1}$ which is the grain boundary dominant term of the full model in Equation \ref{eq2}. Assuming grains are spherical, and since the grain boundaries dominate, this equation can be rewritten as:

\begin{equation} \label{eq3}
    E = \frac{4\pi (\frac{d}{2}^{2}) \times e_{GB} \times \frac{1}{2}}{\frac{4}{3}\pi (\frac{d}{2})^{3}}
\end{equation}
\\

Here, the numerator contains the area of a spherical grain, the energy of grain boundaries per unit area, $e_{GB}$, and the $\frac{1}{2}$ factor as the grain boundary is shared between two grains. The denominator simply contains the volume of the grain. This simply reduces down to:

\begin{equation} \label{eq4}
    E = \frac{3e_{GB}}{d}
\end{equation}
\\

giving, $A = 3e_{GB}$. If we convert this to a per atom value we get:

\begin{equation} \label{eq5}
    A = 3e_{GB}\times atomic\ volume = \frac{3}{2}e_{GB}\ a_{0}^{3}
\end{equation}
\\

A reasonable figure for the energy of grain boundaries, $e_{GB}$, in alpha iron is around $1 - 1.5 \ J/m^{2}$ \cite{SAROCHAWIKASIT2021101186, SHIIHARA2022114268, GBEnergy3}, and when taking $a_{0}=0.286$ nm for alpha iron, the range of $A$ becomes $0.219 - 0.329$ eVnm. Now we assume that the grain are not spherical and instead are cubes. To find an equivalent volume per grain, the following equation is utilised:

\begin{equation} \label{eq6}
    V_{sphere} = \frac{4}{3}pi(\frac{d}{2})^{2} = \frac{1}{6}\pi d^{3} = w^{3}
\end{equation}
\\

where $w$ is the side length of a cube. It follows then that:

\begin{equation} \label{eq7}
   w = \sqrt[3]{\frac{pi}{6}} \ d
\end{equation}
\\

Hence, following the same procedure as above, the value of $E$ becomes:

\begin{equation} \label{eq8}
    E = 3\sqrt[3]{\frac{6}{pi}} \ \frac{1}{d}e_{GB} = 3.72 \ e_{GB} \frac{1}{d}
\end{equation}
\\

Therefore, in the case of the cube grain assumption, the value of $A$ becomes $0.272 - 0.408$ eVnm. Finally, the grains are assumed to be tetrahedrons and an equivalent volume is found compared to the spherical grain approximation. It then follows:

\begin{equation} \label{eq9}
    V_{sphere} = \frac{1}{6}\pi d^{3} = \frac{a^{3}}{6\sqrt{2}}
\end{equation}
\\

where $a$ is the length of the edge. It then follows that:

\begin{equation} \label{eq10}
   a = (\sqrt{2}\pi)^{\frac{1}{3}}d
\end{equation}
\\

As such, we again follow the same procedure as above, and the value of $E$ becomes:

\begin{equation} \label{eq11}
    E = \frac{\sqrt{6} \ 3}{({\sqrt{2}pi})^{\frac{1}{3}}} \ \frac{1}{d}e_{GB} = 4.47 \ e_{GB} \frac{1}{d}
\end{equation}
\\

Hence, when taking an equivalent volume to the spherical approximation, when assuming the grains are tetrahedrons, the value of $A$ becomes $0.326 - 0.489$ eVnm. 

All in all, notwithstanding the approximation of grain shape, the fitted value of $A = 0.307 \pm 8.972\times10^{-4}$ eVnm is coincident with the expected values obtained through an analytical analysis, which gives confidence to the fitting the grain evolution model.

\bibliographystyle{elsarticle-num} 
\bibliography{reference}

\end{document}